\documentclass[epj]{svjour}
\begin{document}
\title{Generalized Fokker-Planck equation: Derivation and exact solutions}
\author{S.I. Denisov\inst{1,2,}\thanks{\rm{e-mail: stdenis@pks.mpg.de}},
Werner Horsthemke\inst{3} \and Peter H\"{a}nggi\inst{4}}
\institute{Max-Planck-Institut f\"{u}r Physik komplexer Systeme, N\"{o}thnitzer
Stra{\ss}e 38, D-01187 Dresden, Germany \and Sumy State University, 2
Rimsky-Korsakov Street, 40007 Sumy, Ukraine \and Department of Chemistry,
Southern Methodist University, Dallas, Texas 75275--0314, USA \and Institut
f\"{u}r Physik, Universit\"{a}t Augsburg, Universit\"{a}tsstra{\ss}e 1, D-86135
Augsburg, Germany}
\date{Received: date / Revised version: date}
\abstract{We derive the generalized Fokker-Planck equation associated with the
Langevin equation (in the Ito sense) for an overdamped particle in an external
potential driven by multiplicative noise with an arbitrary distribution of the
increments of the noise generating process. We explicitly consider this
equation for various specific types of noises, including Poisson white noise
and L\'{e}vy stable noise, and show that it reproduces all Fokker-Planck
equations that are known for these noises. Exact analytical, time-dependent and
stationary solutions of the generalized Fokker-Planck equation are derived and
analyzed in detail for the cases of a linear, a quadratic, and a tailored
potential.
}
\maketitle
\section{Introduction}
\label{Intro} Introduced just 100 years ago \cite{Lang}, the Langevin equation
has become one of the most important and powerful tools for studying noise
phenomena in systems coupled to a fluctuating environment \cite{CKW}. The main
advantage of this equation is that it provides a physically transparent and
mathematically tractable description of the stochastic dynamics of such
systems. The Langevin approach is especially effective if the noise that
describes the action of the environment on the system can be represented as a
time derivative, in the sense of generalized functions, of a stationary process
with {\it independent} increments on non-overlapping intervals. In this case
the solutions of the Langevin equation belong to the class of Markov processes
whose properties are well known (see, e.g., Refs.~\cite{V-K,HL,HT,Risk}). The
stationary processes with independent increments and zero initial state
constitute a class of L\'{e}vy processes \cite{Sato}. For brevity, we call the
L\'{e}vy process, whose derivative produces a given noise, the noise generating
process.

A large variety of physical, biological, financial and other systems is
successfully described by the overdamped Langevin equation, i.e., the
first-order stochastic differential equation. One of the main statistical
characteristics of these systems is the probability density of the solution of
this Langevin equation. If the noise results from the noise
generating process,
then the solution possesses the Markovian property and the probability density
satisfies a closed equation. Sometimes this equation is called the differential
Chapman-Kolmogorov equation \cite{Gard}. On the other hand, it is referred to
as the Fokker-Planck equation for some particular cases. Specifically, a
Gaussian distribution of the increments of the noise generating process
corresponds to the ordinary Fokker-Planck equation \cite{HL,HT,Risk}, and
heavy-tailed stable distributions to the fractional Fokker-Planck equation
\cite{JMF,Dit,MK,YCST,BS,CGKM}. In order to capture these important cases, we
call the closed equation for the probability density that corresponds to an
\textit{arbitrary} distribution of the increments the \textit{generalized}
Fokker-Planck equation. We have shown recently \cite{DHH} that for
\textit{additive} noise the generalized Fokker-Planck equation can be
represented in a unified way through the characteristic function of the noise
generating process. Since the form and coefficients of this equation depend
fundamentally on the distribution of the increments of the noise generating
process, it provides a useful tool for studying the effects of different
noises.

The aim of this paper is twofold. The first is to derive the generalized
Fokker-Planck equation that corresponds to the overdamped Langevin equation
driven by \textit{multiplicative} noise with an \textit{arbitrary} distribution
of the increments of the noise generating process. The second is to solve this
equation for specific cases and, on this basis, to study the distinctive
effects of different noises on the system.

The paper is organized as follows. In Sect. \ref{sec:Leq} we discuss the
overdamped Langevin equation driven by multiplicative noise resulting from the
noise generating process. In Sect. \ref{sec:GFPeq} we derive the generalized
Fokker-Planck equation associated with this Langevin equation in terms of the
transition probability density and characteristic function of the noise
generating process. To confirm the validity of this equation, we consider in
Sect. \ref{sec:SpeCas} various noises, including Poisson white and L\'{e}vy
stable noises, for which the corresponding generalized Fokker-Planck equations
are\linebreak known. Several exact solutions of the generalized Fokker-Planck
equation are derived in Sect. \ref{sec:ExSol}. In Sect. \ref{sec:Concl} we
summarize our results.

\section{Overdamped Langevin equation}
\label{sec:Leq}

The temporal evolution of the relevant degrees of freedom of dynamical systems
that interact with a fluctuating environment is often described by the
(dimensionless) overdamped Langevin equation
\begin{equation}
    \dot{x}(t) = f(x(t),t) + g(x(t),t)\xi(t).
\label{Langevin}
\end{equation}
For different systems the variables in this equation have different meanings,
but to be concrete we will call $x(t)$ [$x(0)=0$] a particle coordinate,
$f(x,t) \!=\! -\partial U(x,t)/\partial x$ a force field, $U(x,t)$ an external
deterministic potential, $\xi(t)$ a random force (noise) resulting from a
fluctuating environment, and $g(x,t)$ a multiplicative noise term. Under
certain conditions (see, e.g., Refs.~\cite{HL,HJ}) the `real' noise $\xi(t)$
with finite correlation time can be approximated by an idealized noise that
effectively captures all the essential noise effects and turns $x(t)$ into a
Markov process, with the result that many of these effects can be described
analytically.

Because of the singular character of the idealized noise, equation
(\ref{Langevin}) has to be interpreted with care. The starting point relies on
the fact \cite{GS} that this noise is the time derivative, in the sense of
generalized functions, of the noise generating process $\eta(t)$. According to
this, the increment $\delta \eta(t) = \eta(t + \tau) - \eta(t)$ of $\eta(t)$ is
\textit{defined} as the time integral, $\delta \eta(t) = \int_{t}^ {t + \tau}
dt'\xi (t')$, in the sense of convergence in distribution. Therefore, the
increment $\delta x(t) = x(t+\tau) - x(t)$ of the particle coordinate during a
time interval $\tau$ ($\tau \to 0$) can be written in the form
\begin{equation}
    \delta x(t) = f(x(t),t)\tau + g(x(t),t)\delta\eta(t),
    \label{incr}
\end{equation}
which defines the meaning of equation (\ref{Langevin}) in the Ito
interpretation \cite{Ito} (see also Refs.~\cite{CKW,HL,HT,Risk}).

For a fixed $\tau$, the distribution of the increments $\delta\eta (j\tau)$
($j=0,1,2,\ldots$) is completely described by the transition probability
density $p(\eta_{j+1}, \tau|\eta_{j})$, where $\eta_{j+1}$ and $\eta_{j}$
denote the possible values of $\eta(j\tau + \tau)$ and $\eta(j\tau)$,
respectively. These densities are properly normalized, $\int_{-\infty}^
{\infty} d\eta_{j+1} \linebreak \times \! p(\eta_{j+1}, \tau| \eta_{j}) = 1$,
and satisfy the following condition $\lim_{\tau \to 0} p(\eta_{j+1}, \tau
|\eta_{j}) = \delta(\Delta \eta)$, where $\delta(\cdot)$ stands for the Dirac
$\delta$ function and $\Delta\eta = \eta_{j+1} - \eta_{j}$. Moreover, if the
first moment of $\eta(j\tau + \tau)$ exists, it is assumed to be zero, i.e.,
$\int_{-\infty}^ {\infty} d\eta_{j+1} \eta_{j+1} p(\eta_{j+1}, \tau| \eta_{j})
= 0$.

The noise generating process, i.e., the stationary Mar\-kov process $\eta(t) =
\lim_{\tau \to 0} \sum_{j=0}^ {[t/\tau]-1} \delta \eta(j\tau)$ with $\eta(0) =
0$ and $[t/ \tau]$ denoting the integer part of $t/\tau$, is also completely
defined by the transition probability density\linebreak $p(\eta_{j+1},
\tau|\eta_{j})$. We note in this regard that all transition probability
densities of the form $p(\eta_{j+l}, l\tau| \eta_{j})$ ($l=2,3,\ldots$) can be
expressed through $p(\eta_{j+1}, \tau|\eta_{j})$ by using the
Chap\-man-Kolmogorov equation \cite{V-K,HL,HT}. In particular, for $l=2$ it
yields $p(\eta_{j+2}, 2\tau|\eta_{j}) = \int_{-\infty} ^{\infty} d\eta_{j+1}
p(\eta_{j+2}, \tau| \eta_{j+1}) p(\eta_{j+1},\linebreak \tau|\eta_ {j})$. Thus,
the statistical properties of solutions of the Langevin equation
(\ref{Langevin}) can be characterized by $p(\eta_{j+1}, \tau|\eta_{j})$ as
well. Next, for simplicity, we additionally assume that $p(\eta_ {j+1}, \tau|
\eta_{j}) = p(\Delta\eta, \tau)$.

If, for example, the transition probability density is Gaussian, i.e.,
\begin{equation}
    p(\Delta\eta, \tau) = \frac{e^{-\Delta\eta ^{2}/(4D \tau)}}
    {\sqrt{4\pi D \tau}},
    \label{trW}
\end{equation}
then $\eta(n\tau) = \sum_{j=0}^{n-1} \delta\eta(j \tau)$ is a discrete-time
Wiener process, which is fully characterized by two parameters,
\begin{equation}
    \langle \delta\eta(j\tau) \rangle = 0, \quad
    \langle \delta\eta(j\tau)\, \delta\eta(l\tau) \rangle =
    2D\delta_{jl}\tau.
    \label{corr}
\end{equation}
The angular brackets denote averaging over the increments $\delta \eta(j\tau)$,
and $\delta_{nm}$ is the Kronecker symbol. These formulas are the discrete-time
versions of the mean $\langle \xi(t) \rangle = 0$ and the correlation function
$\langle \xi(t) \xi(t') \rangle = 2D\delta(t - t')$ of a Gaussian white noise
$\xi(t)$ of intensity $D$. Thus equations (\ref{incr}) and (\ref{corr})
completely specify the Langevin equation (\ref{Langevin}) dri\-ven by
multiplicative Gaussian white noise \cite{HL}.

To derive the Fokker-Planck equation, it is typically assumed that the first
two moments of $\delta\eta(j\tau)$ exist \cite{V-K,Risk}. However, if $p(\Delta
\eta, \tau)$ is a heavy-tailed function of $\Delta \eta$, then the second
moment does not exist. In this case, the derivation of the Fokker-Planck
equation that corresponds to the Langevin equation (\ref{Langevin}) must be
based solely on equation (\ref{incr}). We emphasize that noises characterized
by heavy-tailed transition probability densities $p(\Delta\eta, \tau)$ differ
qualitatively from those characterized by $p(\Delta\eta, \tau)$ with finite
variances. Specifically, the latter have a fre\-quen\-cy independent power
spectral density $\int_{-\infty}^{\infty}dt\, e^{- i\omega t} \langle \xi(0)
\xi(t) \rangle$ ($\omega$ is the frequency) and consequently they
are called white noises. In contrast, the power spectrum of the former does not
exist. Nevertheless, they are a very useful tool for studying an important
class of random processes that exhibit rare but large jumps.

\section{Generalized Fokker-Planck equation}
\label{sec:GFPeq}

We define the probability density of the particle coordinate $x(t)$ in the
usual way:
\begin{equation}
    P(x,t) = \langle \delta(x - x(t)) \rangle.
    \label{defP}
\end{equation}
To derive the evolution equation for this probability density, we need to be
able to express the average values of $F(x(t))$ and $F(x(t), \delta \eta(t))$
in terms of $P(x,t)$; the functions $F(x)$ and $F(x,y)$ are assumed to be
deterministic. Keeping in mind the above definition of averaging, $\langle
F(x(t)) \rangle$ means averaging $F(x(t))$ over all increments\linebreak
$\delta\eta(j\tau)$ with $j = 0,1,\ldots, [t/\tau]-1$ and $\tau \to 0$. It is
obvious from equations (\ref{incr}) and (\ref{defP}) that the result can be
represented as an average with respect to the distribution of $x(t)$, i.e.,
\begin{equation}
    \langle F(x(t)) \rangle = \int_ {-\infty}^{\infty}dx\,F(x)P(x,t).
    \label{F1}
\end{equation}
In order to express $\langle F(x(t),\delta\eta(t)) \rangle$ in terms of
$P(x,t)$, we use a two-stage averaging procedure \cite{DVH}. Since the
variables $x(t)$ and $\delta\eta (t)$ are statistically independent and
distributed according to the probability densities $P(x,t)$ and
$p(\Delta\eta,\tau)$, respectively, we readily obtain
\begin{equation}
    \langle F(x(t),\delta\eta(t)) \rangle = \int_ {-\infty}^{\infty}dx\,
    P(x,t)\int_{-\infty}^{\infty} dy\, F(x,y) p(y,\tau).
    \label{F2}
\end{equation}

To proceed, we introduce the Fourier transform, $P_{k}(t)$, of $P(x,t)$
according to the formula
\begin{equation}
    \mathcal{F}\{ u(x) \} \equiv u_{k} = \int_ {-\infty}^{\infty}
    dx\,e^{-ik x}u(x),
    \label{Ftr1}
\end{equation}
and find, using definition (\ref{defP}), that $P_{k}(t) = \langle e^{-ik x(t)}
\rangle$. Equation (\ref{incr}) implies that for $\tau \to 0$ the increment of
$P_{k}(t)$, i.e., $\delta P_{k} = P_{k}(t+\tau) - P_{k} (t)$,  can be written
in the form
\begin{eqnarray}
    \delta P_{k} &=& -ik\tau\langle e^{-ik x(t)} f(x(t),t) \rangle
    \nonumber\\[6pt]
    && + \langle e^{-ik x(t)}(e^{-ik g(x(t),t)\delta \eta(t)} - 1)
    \rangle.
    \label{incrP}
\end{eqnarray}
In accordance with (\ref{F1}) and (\ref{Ftr1}), the first term on the
right-hand side of equation (\ref{incrP}) reduces to
\begin{equation}
    ik \langle e^{-ik x(t)} f(x(t),t) \rangle =
    \mathcal{F} \bigg\{\frac{\partial}{\partial x}
    f(x,t)P(x,t) \bigg\},
    \label{rel1}
\end{equation}
and equation (\ref{F2}) for the second term gives
\begin{eqnarray}
    &&\langle e^{-ik x(t)}( e^{-ik g(x(t),t)\delta \eta(t)} - 1) \rangle
    \nonumber\\[6pt]
    && = \int_{-\infty}^{\infty} dy\, e^{-ik y}[p_{kg(y,t)}(\tau) - 1]P(y,t),
    \label{rel2}
\end{eqnarray}
where $p_{k}(\tau) = \mathcal{F}\{ p(x,\tau) \} = \langle e^{-ik \delta
\eta(t)} \rangle$ is the characteristic function of $\delta\eta(t)$.

Substituting (\ref{rel1}) and (\ref{rel2}) into equation (\ref{incrP}),
dividing it by $\tau$ and taking the limit $\tau \to 0$, we obtain the
following equation:
\begin{eqnarray}
    \frac{\partial}{\partial t}P_{k}(t) &=& -\mathcal{F}
    \bigg\{ \frac{\partial} {\partial x}f(x,t)P(x,t)\bigg\}
    \nonumber\\[6pt]
    && +\int_{-\infty}^{\infty} dy\, e^{-ik
    y}\phi_{kg(y,t)}P(y,t)
    \label{FPsp1}
\end{eqnarray}
with
\begin{equation}
    \phi_{k} = \lim_{\tau \to 0} \frac{1}{\tau}[p_{k}(\tau) - 1].
    \label{def1}
\end{equation}
Since the transition probability density $p(\Delta\eta, \tau)$ is normalized,
i.e., $p_{0}(\tau) = 1$, the limit (\ref{def1}) must satisfy the condition
$\phi_{0} = 0$. If $k\neq0$, then there exist three different cases, depending
on how quickly $p_{k}(\tau) - 1$ tends to zero as $\tau \to 0$. First, if
$p_{k}(\tau) - 1 = o(\tau)$, then $\phi_{k} = 0$ and the noise does not effect
the system at all. Second, if $p_{k}(\tau) - 1$ tends to zero slower than
$\tau$, then $|\phi_{k}| = \infty$, i.e., the influence of the noise is so
strong that the system relaxes instantaneously to the final state. Finally, the
case we are interested in corresponds to $p_{k}(\tau) - 1 = O(\tau)$, i.e., $0
< |\phi_{\kappa}| < \infty$ and the noise acts on the system in a non-trivial
way.

We apply the inverse Fourier transform, defined as
\begin{equation}
    \mathcal{F}^{-1}\{ u_{k} \} \equiv u(x) = \frac{1}{2\pi}
    \int_{-\infty}^{\infty}dk\, e^{ik x} u_{k},
    \label{Ftr2}
\end{equation}
to equation (\ref{FPsp1}). Using the definition (\ref{def1}), we obtain
\begin{equation}
    \mathcal{F}^{-1}\{ e^{-ik y}\phi_{kg(y,t)} \} = \frac{1}
    {|g(y,t)|} \phi\bigg(\frac{x-y}{g(y,t)}\bigg),
    \label{rel3}
\end{equation}
where the function
\begin{equation}
    \phi(x) = \lim_{\tau \to 0}\frac{1}{\tau}[p(x,\tau) - \delta(x)]
    \label{def2}
\end{equation}
is a special characteristic of $p(\Delta\eta, \tau)$ for $\tau \to 0$ that
describes the influence of noise on the system. Therefore, the desired
generalized Fokker-Planck equation that corresponds to the Langevin equation
(\ref{Langevin}) driven by multiplicative noise, which results from an
arbitrary noise generating process, takes the form
\begin{eqnarray}
    \frac{\partial}{\partial t}P(x,t) &=& -\frac{\partial}
    {\partial x}f(x,t)P(x,t)
    \nonumber\\[6pt]
    && +\int_{-\infty}^{\infty} dy\, \frac{P(y,t)}{|g(y,t)|}
    \phi\bigg(\frac{x-y}{g(y,t)}\bigg).
    \label{FP1}
\end{eqnarray}
In accordance with the definition (\ref{defP}), the solution of this equation
must be normalized and satisfy the initial condition $P(x,0)=\delta(x)$.

To gain more insight into the connection between the generalized Fokker-Planck
equation and the properties of the noise, we introduce the characteristic
function $S_{k} = \langle e^{-ik\eta(1)} \rangle$ of the noise generating
process $\eta(t)$ at $t=1$.  With the formula $\eta(1) = \lim_{\tau \to 0}
\sum_{j=0}^ {[1/\tau]-1} \delta \eta(j\tau)$, it can be rewritten as $S_{k} =
\lim_{\tau \to 0} (p_{k} (\tau))^ {[1/\tau]}$. Then replacing $p_{k}(\tau)$ by
$1 + \tau \phi_{k}$ and taking into account that $\lim_ {\varepsilon \to 0} (1
+ \varepsilon)^{1/\varepsilon} = e$, we find $S_{k} = e^{\phi_ {k}}$, i.e.,
$\phi_ {k} = \ln S_{k}$. Thus, from equation (\ref{FPsp1}) we obtain an
alternative representation of the generalized Fokker-Planck equation:
\begin{eqnarray}
    \frac{\partial}{\partial t}P(x,t) &=& - \frac{\partial}
    {\partial x}f(x,t)P(x,t)
    \nonumber\\[6pt]
    && +\mathcal{F}^{-1} \bigg\{ \int_{-\infty}^{\infty} dy\,
    e^{-ik y}P(y,t)\ln S_{kg(y,t)}\bigg\}.
    \nonumber\\
    \label{FP1b}
\end{eqnarray}
In the particular case of additive noise, where $g(x,t) = 1$, equation
(\ref{FPsp1}) becomes
\begin{equation}
    \frac{\partial}{\partial t}P_{k}(t) = -\mathcal{F}\bigg\{
    \frac{\partial}{\partial x}f(x,t)P(x,t)\bigg\} +
    P_{k}(t) \phi_{k},
    \label{FPsp2}
\end{equation}
and the generalized Fokker-Planck equation (\ref{FP1b}) simplifies to the
equation
\begin{equation}
    \frac{\partial}{\partial t}P(x,t) = -\frac{\partial}{\partial x}
    f(x,t)P(x,t) + \mathcal{F}^{-1}\{P_{k}(t) \ln S_{k}\},
    \label{FP2}
\end{equation}
which was derived in Ref.~\cite{DHH}.

We note that the problem of deriving the generalized Fokker-Planck equation
that corresponds to the Langevin equation (\ref{Langevin}) has been considered
earlier in terms of the L\'{e}vy measure of the noise generating process
\cite{EK1,EK2,DS,DSU}. In contrast, the generalized Fokker-Planck equations
(\ref{FP1b}) and (\ref{FP2}) are derived here in terms of the characteristic
function $S_{k}$ of this process at $t=1$. Since the stationary processes with
independent increments are infinitely divisible, $\ln S_{k}$ can be represented
by the L\'{e}vy-Khin\-tchine formula that connects $S_{k}$ with the L\'{e}vy
measure \cite{Sato}. Hence, both approaches are, in fact, equivalent and lead
to different forms of the generalized Fokker-Planck (see also Sect.
\ref{sec:InfDiv}). It seems, however, that the above approach which deals with
the transition probability density of the noise generating process is more
convenient for applications. Indeed, the transition probability density
completely describes the noise generating process and, in accordance with
(\ref{def1}) and $S_{k}=e^{\phi_{k}}$, explicitly represents the characteristic
function $S_{k}$. At the same time, there is no simple way to invert the
L\'{e}vy-Khin\-tchine formula, i.e., to express the L\'{e}vy measure through
$S_{k}$ \cite{Feller}.

\section{Special cases of the generalized Fokker-Planck equation}
\label{sec:SpeCas}

In order to confirm the validity of equations (\ref{FP1}), (\ref{FP1b}) and
(\ref{FP2}) and to demonstrate their usefulness, we consider several specific
noises for which the corresponding Fokker-Planck equations associated with the
Langevin equation (\ref{Langevin}) are already known.

\subsection{Gaussian white noise}
\label{sec:Gaussian}

The transition probability density $p(\Delta \eta, \tau)$ for Gaussian white
noise is given by formula (\ref{trW}). Accordingly, $p_{k}(\tau) = e^{- D \tau
k^{2}}$, $\phi_{k} = -D k^{2}$, and $S_{k} = e^{-D k^2}$. Then, taking into
account that $\mathcal{F}^{-1}\{ P_{k}(t) k^{2}\} = -\partial^{2} P(x,t)/
\partial x^{2}$, we find that in the case of additive Gaussian white noise
equation (\ref{FP2})  reduces indeed to the ordinary Fokker-Planck
equation \cite{HL,HT,Risk}
\begin{equation}
    \frac{\partial}{\partial t}P(x,t) = -\frac{\partial}{\partial
    x}f(x,t)P(x,t) + D \frac{\partial^{2}}{\partial x^{2}}P(x,t).
    \label{FPGaus1}
\end{equation}

If the Gaussian white noise is multiplicative, then\linebreak $\phi_{kg(y,t)} =
-D k^{2} g^{2}(y,t)$ and
\begin{eqnarray}
    &&\mathcal{F}^{-1} \bigg\{\int_{-\infty}^{\infty} dy\,
    e^{-ik y}\phi_{kg(y,t)}P(y,t)\bigg\}
    \nonumber\\[6pt]
    && = -D \mathcal{F}^{-1}\{ k^{2}\mathcal{F}\{g^{2}(x,t)P(x,t)\}\}
    \nonumber\\[6pt]
    && = D\frac{\partial^{2}}{\partial x^{2}} g^{2}(x,t)P(x,t).
    \label{rel4}
\end{eqnarray}
Applying the inverse Fourier transform to equation (\ref{FPsp1}) and using the
above result, we again obtain the ordinary Fokker-Planck equation
\cite{HL,HT,Risk}
\begin{equation}
    \frac{\partial}{\partial t}P(x,t) = -\frac{\partial}{\partial x}
    f(x,t)P(x,t) + D \frac{\partial^{2}}{\partial x^{2}}g^{2}(x,t)P(x,t),
    \label{FPGaus2}
\end{equation}
which corresponds to the Ito interpretation of the Lan\-gevin equation
(\ref{Langevin}) driven by multiplicative Gaussian white noise. We note that
the last equation can also be derived from equation (\ref{FP1}) with
$\phi(\cdot) = D\partial^{2}\delta (\cdot)/\partial x^{2}$ for this case.

\subsection{Poisson white noise}
\label{sec:Poisson}

As a second example we consider Poisson white noise, i.e., a random sequence of
$\delta$-pulses, defined as \cite{Han1,Han2}
\begin{equation}
    \xi(t) =  \sum_{i=1}^{n(t)}z_{i}\delta(t - t_{i}).
    \label{defPuas}
\end{equation}
Here $n(t)$ is a Poisson counting process with the probability $P(n(t)=n) =
(\lambda t)^{n} e^{-\lambda t}/n!$ of $n \geq 0$ arrivals in the interval
$(0,t]$, $\lambda$ is the rate of the process, $t_{i}$ are the (random) arrival
times of this process, and $z_{i}$ are independent random variables of zero
mean distributed with the same probability density $q(z)$. It is assumed also
that $\xi(t) = 0$ if $n(t) = 0$. The noise generating process $\eta(t)$ is a
step-wise constant Markov process whose increments $\delta \eta(t) = \int_{t}^
{t + \tau} dt'\xi (t')$ are given by
\begin{equation}
    \delta\eta(t) = \left\{ \begin{array}{ll} 0, \quad \textrm{if} \;
    n(\tau) = 0,
    \\ [6pt]
    \sum_{i=1}^{n(\tau)}z_{i}, \quad \textrm{if} \; n(\tau) \geq 1.
    \end{array}
    \right.
    \label{incrPas}
\end{equation}

In order to find the transition probability density\linebreak $p(\Delta \eta,
\tau)$, we use the definition $p(\Delta \eta, \tau) = \langle \delta(\Delta\eta
- \delta \eta(t)) \rangle$ which, together with (\ref{incrPas}), yields
\begin{equation}
    p(\Delta \eta, \tau) = P_{0}(\tau)\, \delta(\Delta\eta) +
    W(\Delta \eta, \tau).
    \label{trans2}
\end{equation}
The first term on the right-hand side of this formula is the probability
density of $\delta\eta(t)$ under the condition that none of the $\delta$-pulses
occurred during the time interval $\tau$. The second term,
\begin{eqnarray}
    W(\Delta \eta, \tau) &=& \sum_{n=1}^{\infty} P_{n}(\tau)
    \int_{-\infty}^{\infty}...\int_{-\infty}^{\infty}
    \delta\bigg(\Delta\eta - \sum_{i=1}^{n} z_{i}\bigg)
    \nonumber\\[6pt]
    && \times \prod_{j=1}^{n} q(z_{j})\,dz_{j},
    \label{defW}
\end{eqnarray}
represents the probability density of $\delta\eta(t)$ under the condition that
at least one pulse occurs during this time interval. Taking the probabilities
$P_{n}(\tau) = P(n(\tau)=n)$ with linear accuracy in $\tau$, i.e., $P_{0}(\tau)
= 1 - \lambda \tau $, $P_{1}(\tau) = \lambda \tau $, and $P_{n\geq 2}(\tau) =
0$,  we obtain from formulas (\ref{trans2}) and (\ref{defW})
\begin{equation}
    p(\Delta \eta, \tau) = (1 - \lambda \tau)\,\delta(\Delta \eta) +
    \lambda\tau q(\Delta \eta).
    \label{trans3}
\end{equation}

In accordance with the definition (\ref{def2}), for this probability density
$\phi(x) = \lambda[q(x) - \delta(x)]$, and the generalized Fokker-Planck
equation (\ref{FP1}) reads
\begin{eqnarray}
    \frac{\partial}{\partial t}P(x,t) &=&
    -\frac{\partial}{\partial x} f(x,t)P(x,t) - \lambda P(x,t)
    \nonumber\\[6pt]
    && + \lambda \int_{-\infty}^{\infty} dy\, \frac{P(y,t)}
    {|g(y,t)|}\, q\bigg(\frac{x-y}{g(y,t)}\bigg).
    \label{FPPois}
\end{eqnarray}
For $g(x,t) = 1$, i.e., in the case of additive Poisson white noise, this
equation is consistent with those reported previously
\cite{Han1,Han2,Kamp,Han3}. Of course, since $S_{k} = e^{-\lambda (1 -
q_k)}$,\linebreak $q(x) = \mathcal{F}^{-1}\{ q_{k} \}$ and $\delta(x) =
\mathcal{F}^{-1}\{ 1 \}$, the same form of the equation (\ref{FPPois}) (with
$g(x,t) = 1$) follows from equation (\ref{FP2}) as well. We note that a wide
class of white noises, which are represented by random sequences of
$\delta$-pulses with a mean number $\lambda$ of pulses per unit time, is
characterized by the same transition probability density (\ref{trans3}).
Therefore, the generalized Fokker-Planck equation (\ref{FPPois}) is also valid
for all these noises.

\subsection{Compound noise}
\label{sec:Compound}

Next we consider the noise $\xi(t) = \sum_{m=1}^{M}\xi_{m}(t)$ composed of a
set of independent noises $\xi_{m}(t)$. In this case the noise generating
process can be written in the form
\begin{equation}
    \eta(t) = \lim_{\tau \to 0}\sum_{m=1}^{M} \sum_{j=0}^{[t/\tau]-1}
    \delta\eta_{m}(j\tau).
    \label{comb}
\end{equation}
Because of the statistical independence of the increments $\delta\eta_{m}
(j\tau)$ of the partial generating processes $\eta_{m}(t)$, the characteristic
function $S_{k}=\langle e^{-ik\eta(1)} \rangle$ of $\eta(1)$ is expressed
through the characteristic functions $S_{mk}=\langle e^{-ik \eta_{m}(1)}
\rangle$ of $\eta_{m}(1)$ as follows: $S_{k} = \prod_{m=1}^{M} S_{mk}$.
Therefore, in the case of additive compound noise the generalized Fokker-Planck
equation (\ref{FP2}) becomes
\begin{eqnarray}
    \frac{\partial}{\partial t}P(x,t) &=& -\frac{\partial}
    {\partial x}f(x,t)P(x,t)
    \nonumber\\[6pt]
    && + \sum_{m=1}^{M} \mathcal{F}^{-1}\{P_{k}(t) \ln S_{mk}\}.
    \label{FP6}
\end{eqnarray}
In particular, if $M=2$ and $\xi_{1}(t)$ and $\xi_{2}(t)$ are Gaussian and
Poisson white noises, respectively, then equation (\ref{FP6}) reduces to
\cite{Gard}
\begin{eqnarray}
    \frac{\partial}{\partial t}P(x,t) &=& -\frac{\partial}
    {\partial x}f(x,t)P(x,t) + D \frac{\partial^{2}}{\partial x^{2}}P(x,t)
    \nonumber\\[6pt]
    && - \lambda P(x,t) + \lambda \int_{-\infty}^{\infty}dy
    P(y,t)\, q(x-y).\qquad
    \label{FP7}
\end{eqnarray}

\subsection{L\'{e}vy stable noise}
\label{sec:Levy}

The generalized central limit theorem \cite{GK} implies that for a wide class
of properly scaled transition probability densities $p(\Delta\eta, \tau)$, the
characteristic function $S_{k}$ corresponds to L\'{e}vy stable distributions,
$S_{k}=S_{k} (\alpha, \beta, \gamma, \rho)$. It is well known \cite{Zol} that
$S_{k}(\alpha, \beta, \gamma, \rho)$ depends on four parameters: an index of
stability $\alpha \in (0,2]$, a skewness parameter $\beta \in [-1,1]$, a scale
parameter $\gamma \in (0, \infty)$, and a location parameter $\rho \in
(-\infty, \infty)$. Assuming in accordance with the initial condition $P(x,0) =
\delta(x)$ that $\rho =0$ and excluding from consideration the singular case
when $\alpha = 1$ and $\beta \neq 0$ simultaneously (in this case $|\phi_{k}| =
\infty$), we obtain $S_{k} = S_{k} (\alpha, \beta, \gamma)$, where \cite{Zol}
\begin{equation}
    S_{k}(\alpha, \beta, \gamma) = \exp\left[ -\gamma |k|^{\alpha}
    \left(1 + i\beta\, \textrm{sgn}(k) \tan \frac{\pi\alpha}{2}\right)\right].
    \label{charfunct}
\end{equation}

In the following we assume for simplicity that the condition $g(y,t) > 0$ holds
for all $y$ and $t$. In this case
\begin{equation}
    \ln S_{kg(y,t)} = g^{\alpha}(y,t)\ln S_{k}(\alpha,\beta,\gamma),
    \label{rel5}
\end{equation}
and the generalized Fokker-Planck equation (\ref{FP1b}) becomes
\begin{equation}
    \frac{\partial}{\partial t}P(x,t)\! = \! -\frac{\partial}{\partial x}
    f(x,t)P(x,t) + \mathcal{F}^{-1}\!\{G_{k}(t)
    \ln \!S_{k}(\alpha,\beta,\gamma)\},
    \label{FP3}
\end{equation}
where
\begin{equation}
    G_{k}(t) = \mathcal{F}\{ g^{\alpha}(x,t)P(x,t) \}.
    \label{rel6}
\end{equation}

Equation (\ref{FP3}) can be rewritten in a form containing the
Riemann-Liouville derivatives defined as \cite{SKM}
\begin{equation}
    _{s}D_{\pm}^{\sigma}h(x) = \frac{(\pm 1)^n}{\Gamma(n - \sigma)}
    \frac{d^n}{d x^n} \int_{0}^{s \pm x} dy\, h(x \mp y)\,
    y^{n - \sigma -1},
    \label{R-L}
\end{equation}
where $_{s}D_{+}^ {\sigma}$ and $_{s}D_{-}^{\sigma}$ denote the operators of
the left- and right-hand side derivatives of the order $\sigma$ ($0<\sigma <
\infty$), respectively. The function $h(x)$ is defined on the interval
$[-s,s]$, $n = 1 + [\sigma]$, and $\Gamma(z)$ is the Gamma function. Using the
characteristic function (\ref{charfunct}), we first represent its natural
logarithm as follows:
\begin{equation}
    \ln S_{k}(\alpha,\beta,\gamma) = -\gamma \frac{(1 + \beta)
    (ik)^{\alpha} + (1 - \beta) (-ik)^{\alpha}}{2\cos(\pi\alpha/2)}.
    \label{rel7}
\end{equation}
Taking the Fourier transform of equation (\ref{R-L}) with $h(x) = g^{\alpha}
(x,t)P(x,t)$, we find
\begin{equation}
    (\pm ik)^{\alpha}G_{k}(t) =
    \mathcal{F} \{_{\infty}D_{\pm}^{ \alpha}g^{\alpha}(x,t)P(x,t)\},
    \label{rel8}
\end{equation}
and combining this result with (\ref{rel7}) we obtain the fractional
Fokker-Planck equation
\begin{eqnarray}
    \frac{\partial}{\partial t}P(x,t) &=& -\frac{\partial}
    {\partial x}f(x,t)P(x,t) - \frac{\gamma}{2\cos(\pi\alpha/2)}
    [(1 + \beta)
    \nonumber\\[6pt]
    && \times _{\infty}D_{+}^{\alpha} + (1 - \beta)_{\infty}
    D_{-}^{\alpha}]\,g^{\alpha}(x,t)P(x,t).
    \nonumber\\
    \label{FP4}
\end{eqnarray}

Equation (\ref{FP4}) reproduces all known forms of the fractional Fokker-Planck
equation that corresponds to the Lan\-gevin equation (\ref{Langevin}) driven by
L\'{e}vy stable noise. It can be easily rewritten in a form containing the
Riesz derivative defined as \cite{SKM}
\begin{equation}
    \frac{\partial^{\alpha}}{\partial |x|^{\alpha}}h(x)
    = -\mathcal{F}^{-1} \{ |k|^{\alpha} h_{k} \}.
    \label{Riesz}
\end{equation}
With the help of this definition and the relations
\begin{equation}
    (_{\infty}D_{+}^{ \alpha} +\! _{\infty}D_{-})h(x) = 2\cos
    \frac{\pi\alpha}{2}\mathcal{F}^{-1} \{ |k|^{\alpha}h_{k} \}
    \label{rel9}
\end{equation}
and
\begin{equation}
    (_{\infty}D_{+}^{ \alpha} - \! _{\infty}D_{-})h(x) = 2\sin
    \frac{\pi\alpha}{2}\frac{\partial}{\partial x}\mathcal{F}^{-1}
    \{ |k|^{\alpha-1}h_{k} \},
    \label{rel10}
\end{equation}
which follow directly from the Fourier representation\linebreak $_{\infty}
D_{\pm}^{\alpha} h(x) = \mathcal{F}^{-1} \{(\pm ik)^{\alpha}h_{k}\}$ of the
Riemann-Liouville derivatives, equation (\ref{FP4}) reduces to
\begin{eqnarray}
    \frac{\partial}{\partial t}P(x,t) &=& -\frac{\partial}
    {\partial x}f(x,t)P(x,t) + \gamma \frac{\partial^{\alpha}}
    {\partial |x|^{\alpha}}\,g^{\alpha}(x,t)P(x,t)
    \nonumber\\[6pt]
    && + \, \gamma \beta \tan\frac{\pi\alpha}{2}\,\frac{\partial}
    {\partial x}\, \frac{\partial^{\alpha -1}}
    {\partial |x|^{\alpha-1}}\, g^{\alpha}(x,t)P(x,t).
    \nonumber\\
    \label{FP5}
\end{eqnarray}
It is obvious that for $\alpha = 2$ this equation takes the form of the
ordinary Fokker-Planck equation (\ref{FPGaus2}) with $D=\gamma$. For
$\alpha<2$, various special cases of equation (\ref{FP5}) have been given
previously in Refs.~\cite{JMF,Dit,MK,YCST,BS,CGKM}.

\subsection{Infinite divisibility of the noise generating process}
\label{sec:InfDiv}

It is well known (see, e.g., Refs. \cite{Sato,Feller}) that any stationary
process with independent increments, including the noise generating process, is
infinitely divisible. This means that the condition $S_{k} = \big( S_{k}^{(n)}
\big)^{n}$ with $S_{k}^{(n)}$ being a characteristic function holds for each
positive integer $n$. In this case, $\ln S_{k}$ can be represented in the form
\cite{Feller}
\begin{equation}
    \ln S_{k} = \int_{-\infty}^{\infty} dz\, \rho(z)
    \frac{e^{-ikz} - 1 + ik \sin z}{z^{2}},
    \label{A1}
\end{equation}
where $\rho(z)$ is the density of the L\'{e}vy measure of $\eta(1)$.
Accordingly, with the integral representation $\delta(\cdot) = (1/2\pi)
\linebreak \times \int_{-\infty}^{\infty} dk\, e^{ik(\cdot)}$ of the $\delta$
function, the last term in equation (\ref{FP1b}) can be written as follows:
\begin{eqnarray}
    &&\mathcal{F}^{-1} \bigg\{ \int_{-\infty}^{\infty} dy\,
    e^{-ik y}P(y,t)\ln S_{kg(y,t)}\bigg\}
    \nonumber\\[6pt]
    && = \int_{-\infty}^{\infty}dz\, \frac{\rho(z)}{z^{2}}
    \int_{-\infty}^{\infty}dy \bigg(\delta[x-y-zg(y,t)]
    \nonumber\\[6pt]
    && \phantom{=} - \delta(x-y) + \sin z \frac{\partial}
    {\partial x}g(x,t) \delta(x-y)\bigg)P(y,t).\qquad\quad
    \label{A2}
\end{eqnarray}
Evaluating the integral over $y$ and using the formula
\begin{eqnarray}
    &&\int_{-\infty}^{\infty} dy\, \delta[x-y-zg(y,t)]P(y,t)
    \nonumber\\[6pt]
    && = \sum_{n=0}^{\infty}(-z)^{n} \frac{\partial^{n}}
    {\partial x^{n}} g^{n}(x,t)P(x,t)
    \nonumber\\[6pt]
    && \equiv \exp\bigg( -z \frac{\partial}{\partial x}
    g(x,t) \bigg)P(x,t),
    \label{A3}
\end{eqnarray}
which follows from the Taylor expansion $\delta[x-y-zg(y,t)] = \sum_{n=0}
^{\infty}(-1)^{n} g^{n}(y,t)(\partial/\partial x)^{n} \delta(x-y)$, we obtain
the generalized Fokker-Planck equation
\begin{eqnarray}
    \frac{\partial}{\partial t}P(x,t) &=& - \frac{\partial}
    {\partial x}f(x,t)P(x,t)
    \nonumber\\[6pt]
    && + \int_{-\infty}^{\infty}dz\, \frac{\rho(z)}{z^{2}}
    \bigg[ \exp \bigg( -z\frac{\partial}{\partial x}g(x,t) \bigg)
    \nonumber\\[6pt]
    && - 1 + \sin z \frac{\partial}{\partial x}g(x,t)\bigg]P(x,t),
    \label{A4}
\end{eqnarray}
which was derived in Ref.~\cite{DS} using a functional approach.

\section{Exact solutions of the generalized Fokker-Planck equation}
\label{sec:ExSol}

An important aspect of the generalized Fokker-Planck\linebreak equation
(\ref{FP2}), which corresponds to the Langevin equation (\ref{Langevin}) driven
by additive noise, is that in some cases it can be solved for all
characteristic functions $S_{k}$, i.e., for all noises represented by a time
derivative of the noise generating process. This provides a unique opportunity
to study in detail the effect of different noises on the same system. In Sects.
\ref{sec:LinPot} and \ref{sec:QuadPot}, we consider this problem for overdamped
particles in linear and quadratic potentials, respectively. The stationary
solution of the generalized Fokker-Planck equation (\ref{FP1b}) in the case of
multiplicative L\'{e}vy stable noise and a tailored potential is presented in
Sect.~\ref{sec:TailorPot}.

\subsection{Linear potential}
\label{sec:LinPot}

In this case $U(x,t) = U(x) = -f_{0}x$ ($f_{0}$ is a constant force acting on a
particle) and, since $\mathcal{F} \{\partial f(x,t)P(x,t)/\partial x \} =
ikf_{0}P_{k}(t)$, equation (\ref{FPsp1}) becomes
\begin{equation}
    \frac{\partial}{\partial t}P_{k}(t) = (-ikf_{0} + \phi_{k})P_{k}(t).
    \label{FPlin}
\end{equation}
The solution of this equation, satisfying the initial condition $P_{k}(0) = 1$
that follows from the initial condition $P(x,0)=\delta(x)$, is given by
\begin{equation}
    P_{k}(t) = \exp(-ikf_{0}t + \phi_{k}t).
    \label{sollin1}
\end{equation}
Therefore, using the relation $S_{k} = e^{\phi_{k}}$, we can write the solution
of the generalized Fokker-Planck equation (\ref{FP2}) as
\begin{equation}
    P(x,t) = \mathcal{F}^{-1} \{S_{k} ^{t} \, e^{-ik tf_{0}} \}.
    \label{sollin2}
\end{equation}

In order to study how the behavior of free particles, when $f_{0}=0$, depends
on the character of the noise, we calculate the second moment $\langle x^{2}(t)
\rangle = \int_{-\infty}^{\infty} dx\, x^{2} P(x,t)$ of the particle coordinate
$x(t)$. Using the solution (\ref{sollin2}) and the integral formula
$\delta''(k) = -(1/2\pi) \int_{-\infty}^{\infty} dx\, x^{2} e^{ikx}$, we obtain
\begin{equation}
    \langle x^{2}(t) \rangle = -\frac{d^2}{dk^2}\, S^{t}_{k}\Big|_{k=0}.
    \label{sec1}
\end{equation}
Since in the case of free particles $x(t) = \eta(t)$, we have $S_{k} =
P_{k}(1)$, and the normalization of $P(x,t)$ yields $S_{0}=1$. Moreover, if the
first moment of $x(t)$ exists, it is assumed to be zero, i.e., $dS_{k}/dk
|_{k=0} = 0$. With these conditions, formula (\ref{sec1}) reduces to
\begin{equation}
    \langle x^{2}(t) \rangle = \langle x^{2}(1) \rangle \,t.
    \label{var2}
\end{equation}

If the term $\langle x^{2}(1) \rangle$, which can be represented as\linebreak
$\langle x^{2}(1) \rangle = \lim_{\tau \to 0} \langle \big[\sum_{j=0}^
{[1/\tau]-1}\! \delta \eta(j\tau) \big]^{2} \rangle = \lim_{\tau \to 0} \langle
\delta\eta^{2}(0) \rangle /\tau$ with $\langle \delta\eta^{2}(0) \rangle =
\langle [\eta(\tau) - \eta(0)]^{2} \rangle = \int_{-\infty}^ {\infty}dy \,y^{2}
p(y,\tau)$, exists, i.e., $\langle \delta\eta^{2}(0) \rangle = O(\tau) $, then
the noise is white and leads to normal diffusion of particles. In particular,
$\langle x^{2}(t) \rangle = 2Dt$ for Gaussian white noise, and $\langle
x^{2}(t) \rangle = \lambda\, t \int_{-\infty}^{\infty}dy \,y^{2} q(y)$ for
Poisson white noise. If the transition probability density $p(\Delta\eta,\tau)$
has heavy tails, i.e., $\langle \delta \eta^{2}(0) \rangle = \infty$, then
$P(x,t)$ evolves in such a way that $\langle x^{2}(t) \rangle$ does not exist.
Such behavior is displayed, e.g., by the probability density $P(x,t) = \mathcal
{F}^{-1} \{S_{k}^{t} (\alpha,\beta,\gamma) \}$ of free particles driven by
L\'{e}vy stable noise with $\alpha \in (0,2)$ \cite{JMF}. While correlated
noise induces a variety of different diffusion regimes of free particles
\cite{DH}, noise resulting from the noise generating process can only give rise
to normal diffusive behavior (if $\langle \delta\eta^{2}(0) \rangle <\infty $)
or to non-diffusive behavior, characterized by a probability density $P(x,t)$
with infinite second moment (if $\langle \delta\eta^{2}(0) \rangle =\infty $).

\subsection{Quadratic potential}
\label{sec:QuadPot}

In the case of a quadratic potential, where $U(x,t) = \linebreak U(x)= b
x^{2}/2$ ($b
> 0$), equation (\ref{FPsp2}) takes the form
\begin{equation}
    \frac{\partial}{\partial t}P_{k}(t) + b k\frac{\partial}
    {\partial k}P_{k}(t) = P_{k}(t) \phi_{k}.
    \label{FP0A}
\end{equation}
Its general solution can be obtained via the method of characteristics
\cite{PZM}, for example, and reads
\begin{equation}
    P_{k}(t) = \exp\bigg( \frac{1}{b} \int_{0}^{k}dz\frac{\phi_{z}}{z}
    + c_{1} \bigg)\, \Psi \bigg( \frac{1}{b}\ln|k| + c_{2} - t \bigg),
    \label{solA1}
\end{equation}
where $\Psi(x)$ is an arbitrary function, and $c_{1}$ and $c_{2}$ are constants
of integration. Since $P_{k}(0) = 1$, the solution (\ref{solA1}) yields
\begin{equation}
    \Psi \bigg( \frac{1}{b}\ln|k| + c_{2} \bigg) = \exp\bigg(\!
    -\frac{1}{b} \int_{0}^{k}dz\frac{\phi_{z}}{z} - c_{1} \bigg).
    \label{relA1}
\end{equation}
Replacing $k$ by $ke^{-b t}$ in this relation, we obtain
\begin{eqnarray}
    \Psi \bigg( \frac{1}{b}\ln|k| + c_{2} -t \bigg) &=&
    \exp\bigg(\! -\frac{1}{b} \int_{0}^{ke^{-b t}}\!
    dz\frac{\phi_{z}}{z} - c_{1} \bigg)
    \nonumber\\[6pt]
    &=& \exp\bigg(\! -\frac{1}{b} \int_{0}^{k}dz
    \frac{\phi_{z e^{-b t}}}{z} - c_{1} \bigg),
    \nonumber\\
    \label{relA2}
\end{eqnarray}
and substituting this result into (\ref{solA1}), we find that
\begin{equation}
    P_{k}(t) = \exp\bigg(\! -\frac{1}{b} \int_{0}^{k}dz
    \frac{\phi_{ze^{-b t}} - \phi_{z}}{z} \bigg).
    \label{solA2}
\end{equation}
Finally, given that $\phi_{k} = \ln S_{k}$, the time-dependent solution of
equation (\ref{FP2}), $P(x,t) = \mathcal{F}^{-1}\{ P_{k}(t) \}$, can be
represented in the form
\begin{equation}
    P(x,t) = \mathcal{F}^{-1} \bigg\{ \exp\bigg(-\frac{1}{b}
    \int_{0}^{k}dz\, \frac{1}{z} \ln \frac{S_{ze^{-b t}}}
    {S_{z}}\bigg)\bigg\}.
    \label{solA3}
\end{equation}

We note that $\ln(S_{ze^{-b t}}/S_{z}) \sim -b tz\, d\ln S_{z}/dz$ as $b \to
0$. Therefore
\begin{equation}
    \lim_{b \to 0} \frac{1}{b} \int_{0}^{k}dz\, \frac{1}{z}
    \ln \frac{S_{ze^{-b t}}}{S_{z}} = -\ln S^{t}_{k},
    \label{relA3}
\end{equation}
and in the case of free particles, i.e, $b=0$, the solution (\ref{solA3})
reduces to $P(x,t) = \mathcal{F}^{-1} \{ S_{k} ^{t} \}$. This result is
confirmed by the solution (\ref{sollin2}) with $f_{0} = 0$.

\subsubsection{Poisson white noise}
\label{sec:PoisWhite}

As a first application of the above results we consider Poisson white noise. In
this case $\phi_{k} = -\lambda (1 - q_{k})$, and formula (\ref{solA2}) yields
\begin{equation}
    P_{k}(t) = \exp\bigg(\! -\frac{\lambda}{b} \int_{0}^{k}dz
    \frac{q_{ze^{-b t}} - q_{z}}{z} \bigg).
    \label{solB1}
\end{equation}
Next we assume that the probability density $q(\Delta\eta)$ is exponential,
i.e., $q(\Delta\eta) = (r/2)\,e^{-r|\Delta\eta|}$ with $r>0$. This implies that
$q_{k} = r^{2}(r^{2} + k^{2})^{-1}$,
\begin{eqnarray}
    \int_{0}^{k}dz \frac{q_{ze^{-b t}} - q_{z}}{z} &=&
    \int_{0}^{k}dz\frac{r^{2}(1 - e^{-2b t})z}{(r^{2} + z^{2})
    (r^{2} + z^{2}e^{-2b t})}
    \nonumber\\[6pt]
    &=& \frac{1}{2}\ln \frac{r^{2} + k^{2}}{r^{2} + k^{2}e^{-2b t}},
    \label{relB1}
\end{eqnarray}
and formula (\ref{solB1}) takes the form
\begin{equation}
    P_{k}(t) = \bigg(\frac{r^{2} + k^{2}e^{-2b t}}
    {r^{2} + k^{2}} \bigg)^{\frac{\lambda}{2b}}.
    \label{solB2}
\end{equation}

For $t \to \infty$ the time-dependent solution $P(x,t) = \mathcal{F}^{-1}\{
P_{k}(t) \}$ of the generalized Fokker-Planck equation (\ref{FP2}) tends to the
stationary solution $P_{\rm{st}}(x) = \mathcal{F}^{-1}\{ P_{k}(\infty) \}$.
Using the relation \cite{PBM}
\begin{equation}
    \int_{0}^{\infty}dk \frac{\cos kx}{(r^{2} + k^{2})^{s}} =
    \frac{\sqrt{\pi}}{\Gamma(s)}\bigg( \frac{|x|}{2r} \bigg)^{s-1/2}
    K_{s-1/2}(r|x|),
    \label{relB2}
\end{equation}
where $\rm{Re\ } s >0$ and $K_{l}(x)$ is the modified Bessel function of the
third kind (or Macdonald function) \cite{BE}, we obtain
\begin{equation}
    P_{\rm{st}}(x) = \sqrt{\frac{2}{\pi}}\, \frac{r (r|x|)^{s-1/2}}
    {2^{s}\Gamma(s)}\, K_{s-1/2}(r|x|)
    \label{solB3}
\end{equation}
with $s = \lambda/(2b)$. This is the so-called $K$-distribution, which is one
of the basic distributions describing the statistical properties of scattered
waves \cite{JT}. It is interesting to note that for $s\leq 1/2$ the stationary
probability density exhibits singular behavior near the bottom ($x=0$) of the
potential well: $P_{\rm{st}}(x) \propto 1/|x|^{1-2s}$ if $0<s<1/2$, and
$P_{\rm{st}}(x) \propto -\ln |x|$ if $s=1/2$.

\subsubsection{L\'{e}vy stable noise}
\label{sec:LevyStab}

In this case $S_{k} = S_{k}(\alpha, \beta, \gamma)$, and the characteristic
function (\ref{charfunct}) implies that
\begin{equation}
    \ln \frac{S_{ze^{-b t}}} {S_{z}} = \gamma (1 - e^{-\alpha
    b t}) |z|^{\alpha} \left(1 + i\beta\, \mathrm{sgn}(z)
    \tan \frac{\pi \alpha}{2}\right).
    \label{relC1}
\end{equation}
Using this result and the integral formula
\begin{equation}
    \int_{0}^{k}dz\, \frac{|z|^{\alpha}}{z}
    \left[\! \begin{array}{cc}
    \mathrm{sgn}\, z \\ [3pt]
    1
    \end{array}\!\right]
    = \frac{|k|^{\alpha}}{\alpha}
    \left[\! \begin{array}{cc}
    \mathrm{sgn}\, k \\ [3pt]
    1
    \end{array}\!\right],
    \label{relC2}
\end{equation}
we find that
\begin{eqnarray}
    \int_{0}^{k}dz\, \frac{1}{z} \ln \frac{S_{ze^{-b t}}}{S_{z}}
    &=& \frac{\gamma}{\alpha} (1-e^{-\alpha b t})|k|^{\alpha}
    \nonumber\\[6pt]
    && \times \left(1 + i\beta\, \mathrm{sgn}(k)
    \tan \frac{\pi \alpha}{2}\right), \qquad
    \label{relC3}
\end{eqnarray}
and the solution (\ref{solA3}) reads
\begin{equation}
    P(x,t) = \mathcal{F}^{-1} \bigg\{ S_{k}\bigg(\alpha, \beta,
    \gamma \frac{1 - e^{-\alpha b t}}{\alpha b} \bigg) \bigg\}.
    \label{solC1}
\end{equation}

We note that special cases of this solution were known previously.
Specifically, the time-dependent solution for symmetric L\'{e}vy stable noise
($\beta=0$) was derived in Ref.~\cite{JMF}, and the steady-state solution for
asymmetric L\'{e}vy stable noise in Ref.~\cite{DGS}.

\subsection{Tailored potential}
\label{sec:TailorPot}

As a third example we derive the stationary probability density function
$P_{\rm{st}} (x)$ for overdamped particles interacting with a tailored
potential
\begin{equation}
    U(x) = c\int_{0}^{x} dy\,y g^{\alpha}(y)
    \label{tailored U}
\end{equation}
($c>0$) and driven by multiplicative L\'{e}vy stable noise with $g(x,t) =
g(x)>0$. In this case $f(x,t) = -cxg^{\alpha}(x)$, and equation (\ref{FP3})
reduces to
\begin{equation}
    c \frac{d}{d x} xG(x) + \mathcal{F}^{-1}\{G_{k} \ln S_{k}
    (\alpha,\beta,\gamma)\} = 0,
    \label{statFP1}
\end{equation}
where $G(x) = g^{\alpha}(x) P_{\rm{st}}(x)$. By applying the Fourier transform
to equation (\ref{statFP1}), we arrive at the ordinary differential equation
\begin{equation}
    -ck \frac{d}{d k} G_{k} + G_{k} \ln S_{k}(\alpha,\beta,\gamma) = 0,
    \label{statFP2}
\end{equation}
whose integration yields
\begin{equation}
    \ln \frac{G_{k}}{G_{0}} = \frac{1}{c} \int_{0}^{k}dz
    \frac{1}{z}\ln S_{z}(\alpha,\beta,\gamma).
    \label{sol1}
\end{equation}
Using the definition (\ref{charfunct}) and the integral formula (\ref{relC2}),
we obtain
\begin{equation}
    G_{k} = G_{0} S_{k}(\alpha,\beta,\gamma/\alpha c),
    \label{sol2}
\end{equation}
and so
\begin{equation}
    P_{\rm{st}}(x) = G_{0} g^{-\alpha}(x)\, \mathcal{F}^{-1}
    \{ S_{k}(\alpha,\beta, \gamma/\alpha c)\}.
    \label{sol3}
\end{equation}
To eliminate $G_{0} = \langle g^{\alpha}(x(t))\rangle$ from the solution
(\ref{sol3}), we use the normalization condition $\int_{-\infty} ^{\infty}dx
P_{\rm{st}}(x) = 1$, which yields the desired probability density
\begin{equation}
    P_{\rm{st}}(x) = \frac{g^{-\alpha}(x)\, \mathcal{F}^{-1}\{
    S_{k}(\alpha,\beta,\gamma/\alpha c)\}}{\int_{-\infty}^{\infty}
    dx\,g^{-\alpha}(x)\, \mathcal{F}^{-1}\{ S_{k}(\alpha,\beta,
    \gamma/\alpha c)\}}.
    \label{sol4}
\end{equation}

To the best of our knowledge, the above result is the first stationary solution
of the fractional Fokker-Planck equation (\ref{FP5}), i.e., the generalized
Fokker-Planck equation (\ref{FP1b}) associated with the Langevin equation
(\ref{Langevin}) driven by multiplicative L\'{e}vy stable noise.

\section{Conclusions}
\label{sec:Concl}

We have derived a new form of the generalized Fokker-Planck equation associated
with the Langevin equation for overdamped particles driven by multiplicative
noise which results from the noise generating process whose independent
increments have an arbitrary distribution. The main advantage of this
generalized Fokker-Planck equation is that it accounts for the noise action in
a unified way, namely through the characteristic function of the noise
generating process at dimensionless time $t = 1$. Since the characteristic
function is completely described by the transition probability density of the
generating process, it is this density which ultimately determines the term in
the generalized Fokker-Planck equation that describes the effect of the noise
on the dynamics of the system. We have explicitly demonstrated this fact for
various noises, including the Poisson white noise and the L\'{e}vy stable
noise.

More importantly, we have solved the generalized Fok\-ker-Planck equation in
the cases of linear and quadratic potentials driven by an arbitrary additive
noise. A remarkable feature of these analytical solutions is that they give an
opportunity to examine the effects of different noises on the same system. As
an illustration, we have derived and analyzed in detail the time-dependent and
stationary solutions that correspond to the Poisson white and L\'{e}vy stable
noises. Moreover, we have also presented the analytical solution of the
fractional Fokker-Planck equation that describes the stationary distribution of
overdamped particles in a specific potential driven by multiplicative L\'{e}vy
stable noise.

\section*{Acknowledgments}

S.I.D. acknowledges the support of the EU through Contract No.
MIF1-CT-2006-021533, and P.H. acknowledges financial support by the
Deutsche For\-schungs\-ge\-mein\-schaft via the Collaborative
Research Centre SFB-486, Project No. A 10, and by the German
Excellence Cluster ``Nanosystems Initiative Munich" (NIM).

\end{document}